# Signature of gate tunable superconducting network in twisted bilayer graphene

Yingbo Wang[1,2]†, Yingzhuo Han[1]†, Lu Cao[2]†, Xun-Jiang Luo[3], Yucheng Xue[1], Jiefei Shi[2], Xiaomeng Wang[2], Xiangjia Bai[1], Junnan Jiang[2], Ziyi Tian[2], Kenji Watanabe[4], Takashi Taniguchi[5], Fengcheng Wu[3], Qing-feng Sun[6], Hong-Jun Gao[7], Yuhang Jiang[2]*, Jinhai Mao[1]*

[1] School of Physical Sciences, University of Chinese Academy of Sciences; Beijing, 100049, China.

[2] College of Materials Science and Optoelectronic Technology, Center of Materials Science and Optoelectronics Engineering, University of Chinese Academy of Sciences; Beijing, 100049, China.

[3] School of Physics and Technology, Wuhan University; Wuhan, 430072, China.

[4] Research Center for Electronic and Optical Materials, National Institute for Materials Science; Tsukuba, 305-0044, Japan.

[5] International Center for Materials Nanoarchitectonics, National Institute for Materials Science; Tsukuba, 305-0044, Japan.

[6] International Center for Quantum Materials, School of Physics, Peking University; Beijing, 100871, China.

[7] Beijing National Center for Condensed Matter Physics and Institute of Physics, Chinese Academy of Sciences; Beijing, 100190, China.

*Corresponding author. Email: yuhangjiang@ucas.ac.cn; jhmao@ucas.ac.cn.

†These authors contributed equally to this work.

**Abstract:** Twisted van der Waals materials provide a tunable platform for investigating two-dimensional superconductivity and quantum phases. Using spectra-imaging scanning tunneling microscopy, we study the superconducting states in twisted bilayer graphene and track their evolution from insulating phases. Gate-dependent spectroscopic measurements reveal two distinct regimes: under-doped ($v$ = -2.3) and optimally doped ($v$ = -2.6). In the under-doped regime, partial superconductivity arises, forming a network interspersed with non-gapped regions. At optimal doping, the entire unit cell demonstrates superconductivity, with gap size modulation showing an anti-correlation with the local density of states. This gate-dependent transition from an insulating phase to a modulated superconductor uncovers an unexpected spatial hierarchy in pairing behavior and offers direct microscopic insights to constrain theories of superconductivity in moiré systems.

Doping a correlated insulator with controllable carrier concentration to a superconductivity or metallic regime has been a powerful path for exploring novel quantum phases and understanding the physics of its ground state(*1-8*). This quantum phase transition near its critical points is characterized by significant quantum fluctuations due to the competition between superconducting (SC) order and localization effects, which can observably modify the SC states(*9*). For instance, the finite carrier doping in cuprates commonly leads to the emergence of a randomly distributed pseudo-gap state or a long-range pair density wave(*10-12*), where both are believed as the precursor of high-temperature superconductivity. Recently, exotic superconductivity has been realized in the artificial van der Waals heterostructure after finely controlling the interlayer misalignment angle and carrier doping, such as in twisted bilayer graphene (tBG)(*1, 2*). The high crystalline tBG can eliminate the interference of impurities or inhomogeneity, so that we can study the intrinsic nano-scale spatial evolution of SC state near the phase transition point. This is crucial to reveal the SC pairing mechanism and the relationship between SC and correlated insulating states.

Probe measurements of the local density of states (LDOS) have been widely used in the study of both conventional and unconventional superconductors, providing crucial evidence for understanding the underlying pairing mechanism(*1, 2, 13-17*). In this study, we utilized spectra-imaging scanning tunneling microscope (SI-STM) to investigate the SC state in tBG. Our d$I$/d$V$ spectrum reveals that superconductivity in tBG sustains at a temperature around 6 K, which is higher than previously reported values(*3-5, 18, 19*). To confirm the presence of superconductivity, we perform point contact spectrum (PCS) measurements, temperature dependent spectrum, and magnetic field dependent spectrum. In this work, four tBG samples (named Sample 1, 2, 3 and 4) exhibit superconductivity above 4.4 K. We primarily focus on results from Sample 1, and information of Sample 2, 3 and 4 are provided in Supplementary Text (ST), Section I – III.

One of the key results is that the SC gap in tBG exhibits a moiré period-dependent modulation, rather than being uniformly distributed. Under finite hole doping near $v$ = -2.3, which we classify as the under-doped regime, the AB stacked regions (dark area in Fig. 1b) of the moiré superlattice first transition into the SC phase while the AA stacked regions remain in a normal metallic state. With further doping near $v$ = -2.6 (classified as optimal-doped regime), both AA and AB stacked regions enter the SC phase. However, in this doping regime, a gap modulation emerges with maximum (minimum) gap values and lowest (highest) LDOS located at AB (AA) stacked region, leading to a π phase difference between the SC gap and LDOS modulation. We propose a phenomenological model which could qualitatively explain the experimental results. Our experimental results and theory not only provide valuable insights into the pairing mechanism and evolution of superconductivity, but also establish a possible connection between the 2D moiré system and the heavy fermion system. It is also proved that 2D moiré system is an ideal platform for SC manipulation.

**Electronic structure and correlation effect for tBG at the magic angle**

The electronic structure of tBG is investigated by a scanning tunneling microscope (STM) setup, as sketched in Fig. 1a. The tBG samples are fabricated using the modified 'tear and stack' technique (see Method)(*20*), with their doping level adjustable by a gate voltage ($V_g$) applied to the silicon substrate. The STM topography in Fig. 1b reveals the characteristic moiré superlattice structure in tBG. The bright protrusion in the topography corresponds to AA stacked region, while the dark triangular area represents the AB/BA stacked region with the interposed domain walls (DW) (dashed lines). According to the moiré periodic length measured by $L = a/2\sin(\frac{\theta}{2}) = 12.9$ nm, where $a$ is the lattice constant of graphene, the twist angle $\theta$ can be extracted as 1.09°, which falls within the 'magic angle' region(*21*).

For this magic angle, two weakly-dispersive flat bands appear near the Fermi level ($E_F \equiv 0$ meV in STM experiment), resulting in an enhanced LDOS(*22-24*) (red bands in Fig. 1c, more details in ST, Section IV). Each of these flat bands in tBG has a fourfold degeneracy, accounting for the isospin (spin and valley) degrees of freedom(*25-27*). The filling factor $v$ of the tBG is defined as +4 when the flat bands are completely filled and as -4 when they are entirely empty. After identifying the filling $v = \pm 4$ by examining the energy position of these flat bands relative to $E_F$, other partial fillings could be directly determined (Fig. 1d and 1e, more details in ST, Section V). Additionally, the theoretical calculation has shown that the LDOS for those flat bands primarily concentrates on the AA stacked regions, exhibiting a spatially modulated distribution following the moiré structure(*21, 22, 28*). Our d$I$/d$V$ spectra confirm the similar modulation of LDOS within the moiré unit cell, specifically stronger LDOS for the flat bands in the AA than that in AB stacked regions (more details in ST, Section VI).

The kinetic energy of electrons in these flat bands is significantly reduced compared to the Coulomb interaction strength, leading to the emergence of correlation effects in the electronic

states(*29, 30*). One mechanism for the correlated insulating (CI) states at integer fillings is the interaction-driven isospin flavor polarization and degeneracy lifting(*31-34*). This can be observed in the gate-dependent d$I$/d$V$ spectra, where one isospin flavor moves downward across $E_F$ to lower energy at each integer filling(*31, 35*) (detailed discussion on the filling process in ST, Section VII). In Fig.1d and e, we plot the d$I$/d$V$ intensity as a function of $v$ ($V_g$) and energy ($V_{Bias}$), highlighting the process of isospin flavor lifting with black dashed lines starting from $v$ = -4, -3 and -2. Additionally, CI states are also observed near the filling factors $v$ = +2 and +3 as indicated by the orange arrows(*18, 19, 36, 37*).

**Observation of SC states**

Now we focus on the electronic states of tBG in the doping range -3 < $v$ < -2, where both transport and STM measurements have revealed the presence of a SC phase(*2, 3, 5, 38*). In Fig. 1d and e, we have noticed that a clear gap feature emerges at $E_F$ for partial filling between $v$ = -2 and -3 in both AA and AB stacked regions (indicated by red arrows). To better elucidate this gap feature, Fig. 2a presents the plot of d$I$/d$V$ intensity with higher energy resolution and finer $V_g$. Notably, two peaks are identified at the gap edges (~±3 meV). Similar results are also independently observed in Sample 2, 3 and 4 (ST, Section I – III). To minimize the influence of background effects on the gap analysis, especially from the nearby flat bands, we also use the second derivative of -d$I$/d$V$ in subsequent discussions for clarity. It is important to note that this technique does not affect the energy position of the coherence peaks and is easier to identify the energy position(*13, 39*), as shown in direct comparison in Fig. 2b (other comparisons are in ST, Section VIII). The gap feature flanked by two peaks is also pronounced in -d$^3I$/d$V^3$ at given positions depicted in Fig. 2c and 2d. This gap feature, together with its occurrence at the non-integer filling, hints the emergence of a SC phase in our tBG.

To ensure that the observed gap is indeed induced by the emergent superconductivity, we perform PCS measurements, which have been proved effective in identifying the SC phase from gap features in STM experiments(*1, 2*). During PCS measurement, when the tip approaches the surface of the superconductor, the tunneling junction barrier decreases gradually. As a result, electron from the normal metallic tip at energy less than SC gap enters the superconductor and forms a Cooper pair, accompanied with the retroreflection of a hole (Fig. 2e). This so-called Andreev reflection leads to an enhancement of the intensity in the d$I$/d$V$ spectrum within SC gap(*40, 41*). Figures 2f and 2g show a series of d$I$/d$V$ spectra tracing the transition from normal tunneling to Andreev reflection. As the tip gradually approaches the surface of tBG by increasing the setpoint of tunneling current, there is a progressive increase in the d$I$/d$V$ spectral intensity within the SC gap until it eventually develops into a peak at $E_F$ caused by the Andreev reflection. This confirms the scenario that conductivity induced by Andreev reflection increases with the setpoint of tunneling current. All the observation confirms our assumption that the energy gaps observed near $v$ = -2.3 and -2.6 are attributed to superconductivity.

For a clear contrast, we also performed the PCS measurements on the CI states at $v$ = +2 with a similar gap feature in the normal STS (Fig. 2h). Unlike the emergence of Andreev reflection peak at the setpoint of 35 nA in the SC regime, no evidence of Andreev reflection can be observed for this CI state even with the tunneling current as high as 100 nA or larger (see more PCS without Andreev reflection in ST, Section IX). These comparative analyses provide strong support that the observed gap at $v$ = -2 ~ -3 in magic angle tBG is attributed to the SC state, and effectively rule out the affect from the CI state or other extraneous factors. In addition to PCS measurements, the SC gap is also confirmed by adding a magnetic field perpendicular to the surface (Fig. 2m and 2n). The suppression of the energy gap by the magnetic field is also consistent with a SC state(*4, 18*).

In addition, we conduct the temperature-dependent d$I$/d$V$ spectrum measurement on this SC state from 4.5 K to 9.1 K (Fig. 2j). At 6.0 K, although the two coherence peaks are still weakly observed, the SC gap between them nearly disappears and the minimum of gap deviates significantly from the Fermi level. These spectra have the same characteristics as the spectra of pseudogap previously reported in tBG(*2*). When the temperature reaches 8.0 K, both the coherence peaks and the gap vanish (see ST, Section X for more details). The SC transition temperature in our sample is above 4.4 K, which is slightly higher than the value recorded in transport measurements(*4*). This discrepancy may be attributed to our local measurement approach which minimizes fluctuations and disorder effects. Moreover, we also measure the spectrum at 0.4 K, where the Dynes-formula fitting suggests the similar feature of nodal superconductivity as previously reported (Fig. 2k and 2l, details of fitting in ST, Section XI)(*2*). The intensity of the conductance at zero bias in the AA stacked region is closer to zero, which resembles a d-wave SC behavior. However, the intensity at zero-bias is higher in AB stacked region, which may be caused by the relatively strong scattering in the AB stacked region.

The above discussions confirm the SC state in AA and AB stacked regions near $v$ = -2 ~ -3. Intriguingly, the fittings in Fig. 2k and 2l present a SC-gap-size difference between AA and AB regions (4.10 meV in AA stacked region and 6.25 meV in AB stacked region), a signature of spatially modulated pairing order. And such SC gap size modulation is also obviously reflected in the energy position of the coherence peak (Fig. 2d). In the following discussion, we will define $2\varDelta$ as the SC energy gap between two coherence peaks and explore its spatial evolution.

**Evidence for spatially modulated superconductivity**

With further analyses of the SC states, we find that gap feature occurs at different fillings in AA and AB stacked regions. That is, at $v = -2.3$, the AB regions exhibit superconductivity, while the AA regions keep their gapless feature, indicating a normal metallic state (Fig. 2c and 2i). To ensure clear filling boundary of SC regimes, Fig. 2i shows the filling-dependence PCS measurement from $v = -3.0$ to $-2.3$, which indeed gives different filling regimes of SC states (red dashed rectangles) in AA and AB stacked regions. This unexpected finding suggests that superconductivity has a higher priority in AB stacked regions than AA ones. To facilitate subsequent discussion and differentiation, we define the doping regime near $v = -2.3$ as under-doped regime. Notably, in the similar under-doped cuprate systems, the local SC gaps randomly distributed before global superconductivity forms(*39, 42*), while in tBG the local SC states show a periodic emergence in AB stacked regions due to the well-formed moiré superlattices in real space. The potential ways of forming global superconductivity within this under-doped regime will be discussed later.

After further increasing the filling to the optimal-doped regime ($v = -2.6$), both the AA and AB stacked regions undergo a transition into the SC phase, corresponding to the SC near $v = -2.6$ in transport measurements(*3-5, 18, 19, 36, 43-45*). Strikingly, even in this optimal-doped regime, there remains a distinct SC-gap-size difference between AA and AB stacked regions (4.4 K data in Fig. 2d and 0.4 K data in Fig. 2k-2l). To explore the spatial modulation and evolution of the SC gap in this moiré system, we perform spatially dependent d$I$/d$V$ spectra along one high symmetry direction in the moiré superlattice. Along direction I (AA-AB-DW-BA) in Fig. 3a and 3b, the orange dashed line is utilized to trace the positions of the right coherence peak, which clearly demonstrates the oscillating feature. To better elucidate the oscillation, we plot the second derivative of -d$I$/d$V$ spectra in Fig. 3c and 3d. Additionally, the extracted SC gap size from each spectrum is provided in Fig. 3e (also see coherence peak height modulation in ST, Section XII). The result further proves that the SC gaps in AA stacked regions and DW are smaller than those in AB/BA areas. Moreover, the modulation of SC gap has a periodicity close to the moiré period.

To accurately determine the periodicity, we perform the one-dimensional Fast Fourier transform (FFT) to oscillating gap value in ST, Section XIII. The FFT intensity displays a prominent peak at $k_a = 0.0929$ nm$^{-1}$ corresponding to a periodicity of $L_a = 10.76$ nm. This periodicity closely aligns with the distance between the AA stacked region and the nearest neighbor DW along direction I. We also use the $k_a$ value obtained by FFT to fit the data in Fig. 3e (dashed line) with a sinusoidal function, $2\Delta = |a\sin(\pi k_a x + \varphi)| + c$, where $x$ is the distance, $a$, $\varphi$ and $c$ are the amplitude, phase and constant obtained by fitting (see ST, Section XIII). The fitted curve well reproduces the oscillating behavior of $2\Delta$ and affirms that the SC gap size in AB stacked regions is larger than that in AA stacked regions (Fig. 3e and ST, Section XIV). The spatially dependent spectra along other directions showing the similar gap oscillating phenomena could be found in ST, Sections XV. All the evidence suggests that SC gap size is influenced by the underlying moiré pattern which leads to a spatial modulation of superconductivity in tBG.

We further confirm that this SC gap's modulation in the AA and AB stacked regions exists in a wide doping range. Figure 3f shows the evolution of the SC gap in these regions under different filling factors ($v$ from -2.4 to -2.7). Due to the confirmation of the filling range of SC states by PCS measurements (Fig. 2i), we can focus on $v$ values away from the CI states with similar gap features. Despite the small variation in the SC gap ($2\Delta$) under different fillings, the SC gaps in AB stacked regions (blue data) are generally larger than those in AA regions (red data), indicating universal behaviors within our samples. The more robust superconductivity and larger gap value in the AB stacked regions are also observed in sample 3 and 4, which suggests a common feature for the tBG systems.

With this discovery, we further use the SI-STM to unveil the full distribution of the SC gap in moiré lattice. Figure 4 depicts the maps of extracted SC gap size obtained from spectra acquired at each spatial point (corresponding topography of unit cell in inset of Fig. 4a, more details in ST, Section XVI). Consistent with the findings presented in Fig. 3, these maps also reveal a spatial modulation of SC gap following the periodicity of the moiré superlattice (dashed green rhombus in maps).

Notably, the maps unveil a unique phase transition from a CI state to SC behavior in the tBG system. Figure 4a shows a schematic phase diagram based on the measured results. Firstly, in the under-doped regime, only a portion of each moiré unit cell transitions into the SC state (red shadow) instead of maintaining a globally uniform gap. This global mode of phase transition has similarities with that observed in the under-doped cuprates, where some local SC states first emerge in real space(*39, 42*). However, unlike the disordered modulation in cuprates, the modulation in tBG is highly periodic (Fig. 4b). So, for the under-doped maps ($v = -2.3$), the AB stacked regions firstly exhibit superconductivity, forming a network-like structure to achieve global superconductivity across the entire sample. These experimental images contradict the previous theoretical results, which suggest a higher SC pairing amplitude or order parameter in the AA stacked regions compared to the AB stacked regions, due to the larger LDOS in AA stacked regions(*46-48*). Our experimental results suggest a more robust SC phase in the AB stacked regions. Secondly, in the optimal-doped regime ($v = -2.6$), the entire area of each moiré unit cell transitions into SC phase; however, variations in gap size persist between AA and other regions (Fig. 4c). Figure 4c shows a clear suppression of the SC gap size in the AA stacked regions (much lighter color in the corner of unit cell). Furthermore, the period of this modulation is about 12.85 nm, completely following the periodicity of the moiré superlattice in sample 1(see the SC gap modulation at other fillings in ST, Section XVI). In summary, although the LDOS in the AA stacked regions has a larger distribution, AA stacked regions transition into the SC phase later when the system across under-doped regime, and present a smaller SC gap size when the system in optimal-doped regime. Consequently, an anti-correlation between LDOS distribution and SC gap modulation within moiré superlattices emerges, resulting in a π phase difference between SC pairing order and LDOS concertation.

Now we discuss the potential mechanisms to elucidate the observed experimental anticorrelation between the LDOS and the SC energy gap. The first possibility is the spontaneous formation of a pair density wave(*13, 49*), which represents an exceptional SC state, characterized by the remarkable phenomenon of pairing gap modulation in real space that leads to the translation symmetry breaking. While our SC gap modulation does not break the translation symmetry of the moiré superlattice, its anticorrelation with the LDOS mimics characteristics of pair density wave. Additionally, in the under-doped regime, there exist non-gapped AA stacked regions, which is reminiscent of the theoretically proposed one-dimensional pair density wave in cuprates(*50*). Whether the observed SC modulation could be ascribed to the pair density wave still needs further experimental and theoretical works.

The second possibility is spontaneous charge modulation driven by electron-electron interaction, which modulates the local doping within the moiré unit cell. Similar SC network has been deduced from transport measurement in 1T-TiSe$_2$ and tBG system near the insulator-superconductor transition point(*4, 51*). In those works, either the doping induced insulator-superconductor phase separation or carrier inhomogeneity is believed as the source. Our system features the similar phenomena but has a shorter periodicity. The STM imaging enables us to exclude the charge inhomogeneity scenario, because there is no signature on the local band-filling difference within the moiré unit cell (*e.g.*, the AA and AB stacked regions have different gate voltages corresponding to the same *v*), as we previously reported in tBG or twisted monolayer-bilayer graphene(*52, 53*).

We propose a phenomenological model (details in ST, Section XVII) based on a site-dependent electron-electron interaction to explain spatially modulated SC gaps. This model offers qualitative insights into why SC emerges first in low LDOS regimes. Additionally, we note that a similar spatial dependent SC gap at the atomic scale is observed in heavy Fermi superconductivity(*54, 55*), in which the gap value near the local moment region is smaller than that far away. Our results feature this character, especially considering the recent theoretical proposals that heavy fermion physics emerges at such carrier doping regimes (details in ST, Section XVIII)(*56-58*). However, further experimental and theoretical research efforts are still needed to unveil its mechanism.

In conclusion, our study focuses on the phase transition process from a correlated insulator to superconductivity in tBG. We investigate the SC state between *v* = -2 and -3 using gate-tuned SI-STM/STS and confirm it through PCS analysis. Additionally, we reveal the complex spatial distribution of the gap as it evolves from the CI to the SC regime. The SC behavior in tBG exhibits strong gate tunability and spatial configuration, which appear to surpass current theoretical expectations. Moreover, its spatial distribution features characteristics observed in heavy fermion superconductivity(*54-58*) or cuprate systems with spatially gap modulation(*10, 13, 39*). These findings shed light on uncovering correlated phases in other quantum systems.

**Acknowledgments:** We appreciate the enlightening discussions with Prof. Xi Dai, Jian Kang, Zhi-Da Song and Hai-Hu Wen.

**Author contributions:**

Conceptualization: J.M., Y.J.

Methodology: J.M., Y.J., Y.W., Y.H., L.C., Y.X., J.S., X.W.

Investigation: Y.W., Y.H., L.C.

Visualization: Y.W., Y.H., L.C., X.L.,

Funding acquisition: J.M., Y.J. L.C.

Project administration: J.M., Y.J.

Supervision: J.M., Y.J.

Writing – original draft: J.M., Y.J., Y.W., Y.H., L.C.

Writing – review & editing: All authors read and approved of the final version of the manuscript.

**Competing interests:**

The authors declare no competing interests.

**Data and materials availability:**

All data are available in the main text or the supplementary materials.

**Supplementary Materials**

Materials and Methods

Supplementary Text

Figs. S1 to S19

Tables S1 to S2

References

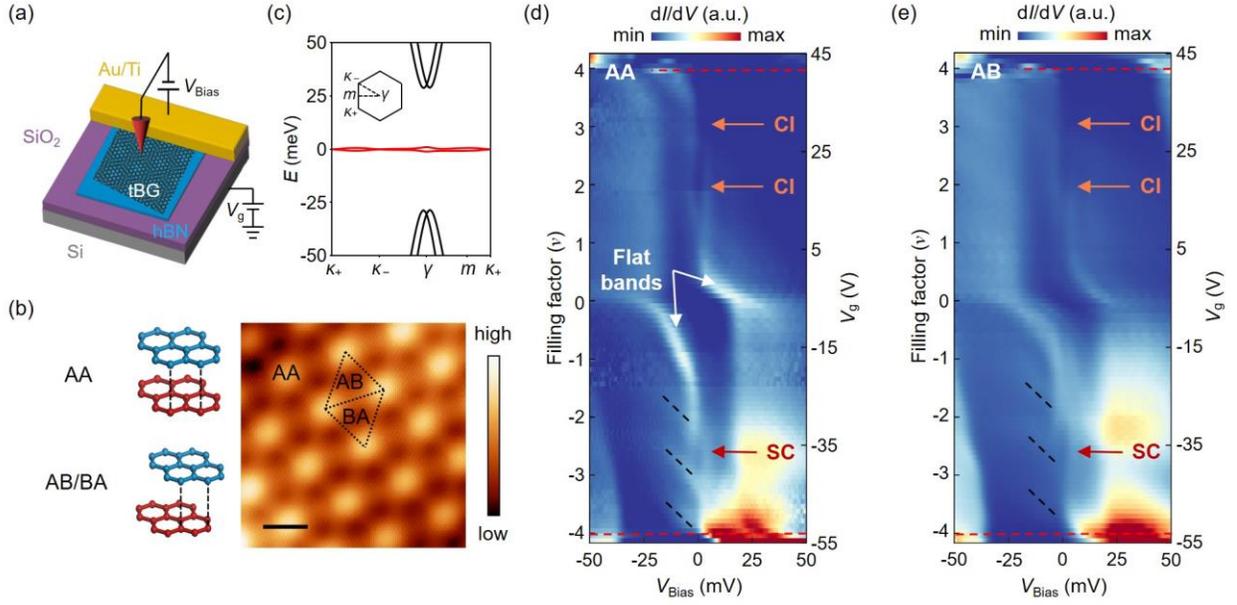

**Fig. 1 Structure of tBG and gate tunable correlated states.** (a) Schematic drawing of the device and STM experimental setup. $V_{Bias}$ represents the bias voltage applied between the sample and tip, and $V_g$ denotes the gate voltage used for carrier concentration tuning. (b) Left: Schematic of AA and AB stacked region; Right: STM topography of the tBG at $V_{Bias}$ = -100 mV and $I$ = 0.3 nA; the scale bar is 10 nm; In this topography, AA stacked regions appearing as bright spots, AB/BA stacked regions as dark areas, and DW connecting AA stacked regions indicated by dashed lines. (c) Band structure calculated using a continuum model for tBG, with red highlighting the two flat bands; the inset provides a sketch of its Brillouin zone. (d-e) Gate-dependent d$I$/d$V$ intensity plot of the AA (d) and AB (e) stacked region ($V_{Bias}$ = -50 mV, $I$ = 0.6 nA). The flat bands, superconducting (SC) regime and correlated insulating (CI) states are indicated by arrows. Black dashed lines highlight the occupied isospin flavor due to the electron-electron interaction driven flavor polarization process at integer filling, while red dashed lines highlight filling factors ±4.

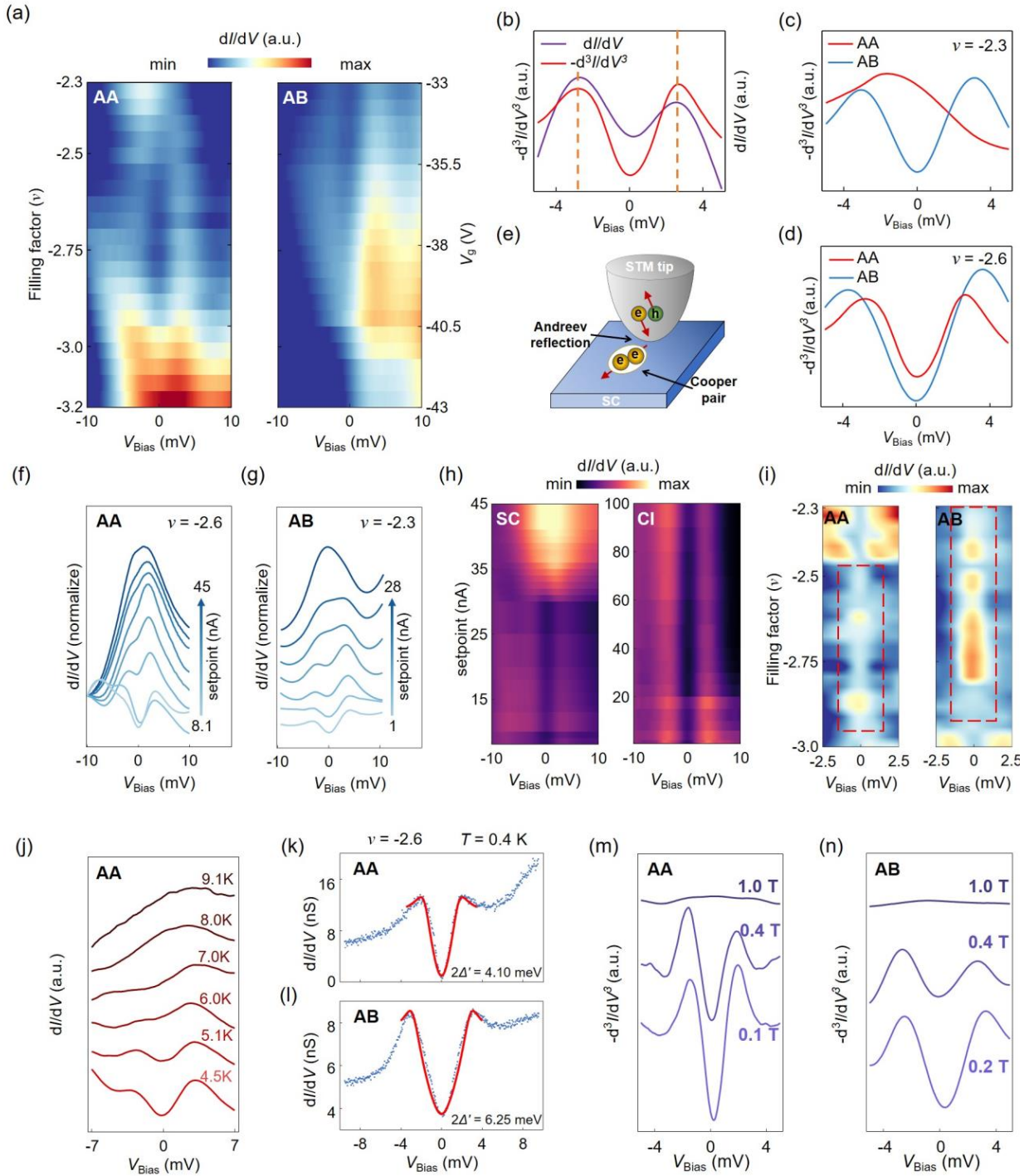

**Fig. 2 Identification of SC states in AA and AB staked regions in electronic doping (v) from -2.3 to -3.0.** (a) High-resolution d$I$/d$V$ intensity plot for the AA and AB stacked regions with $v$ from -3.2 to -2.3. (b) The direct comparison between the original d$I$/d$V$ and -d$^3I$/d$V^3$ spectrum at $v$ = -2.6, with orange dashed lines highlighting coherence peaks. (c-d) The -d$^3I$/d$V^3$ spectra for the AA and AB stacked regions at $v$ = -2.3 (c) and $v$ = -2.6 (d). (e) A schematic drawing illustrates Andreev reflection occurring at the interface between a normal metallic STM tip and an SC sample. (f-g) Point contact spectra (PCS) in SC phase for the AA (f) and AB (g) stacked region. The spectra are normalized for clarity. A larger setpoint value represents a smaller distance between the tip

and the sample. (h) PCS measured in SC phase (left, $v = -2.6$) and CI phase (right, $v = +2.1$). The normalized PCS shows that as the tip approaches gradually towards the sample surface, a peak feature emerges near Fermi level in SC phase from 35 nA while maintaining gap feature in CI state. (i) Gate-dependent PCS of SC phase in the range of $v = -2.3$ to $-3.0$ for the AA (left) and AB (right) stacked region. SC exists only in the AB stacked region at under-doped regime. The red dashed rectangles label the SC phase. (j) Temperature-dependent d$I$/d$V$ spectra for the SC state for AA stacked region recorded at $v = -2.6$. (k-l) Dynes-function fits to the experimental tunnelling spectrum (red curve) measured at $v = -2.6$ for AA (k) and AB (l) stacked region at 0.4 K using the model of quasiparticle DOS for a nodal superconductor. (m-n) Magnetic field dependent -d$^3I$/d$V^3$ spectra for the AA (m) and AB (n) stacked region at $v = -2.6$. The magnetic field is perpendicular to the surface.

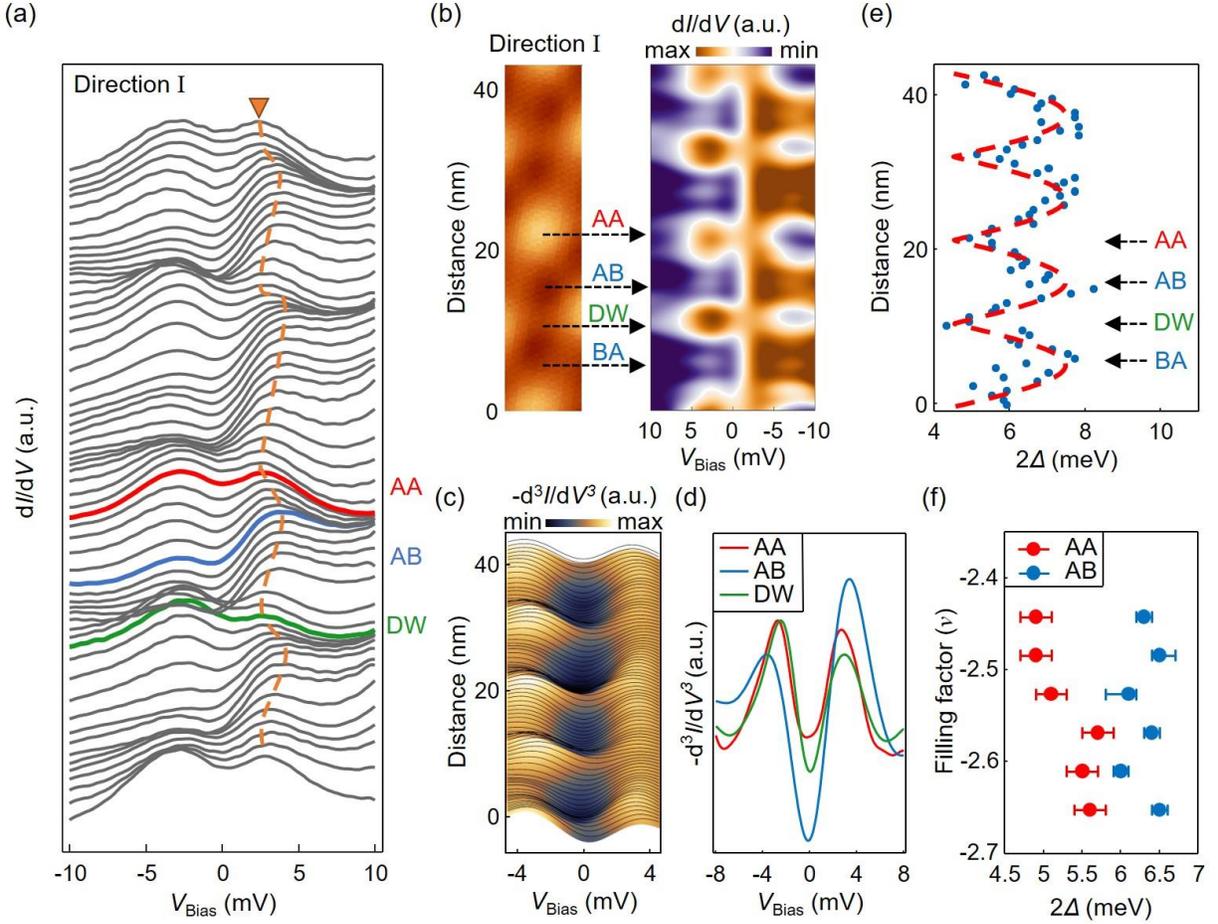

**Fig. 3 Spatially modulated SC spectrum in optimal-doped regime.** (a) The spatial dependent d$I$/d$V$ spectra along the direction I ($V_{Bias}$ = -50 mV, $I$ = 0.6 nA, $v$ = -2.6). Typical spectra at AA, AB, DW stacked regions are highlighted in red, blue, and green, respectively. The orange dashed line is the guide to the eyes about the evolution of right coherence peak with space. (b) The intensity plot of the d$I$/d$V$ spectra (right panel) along the direction I shown in the topography (left panel). Different stacked regions are highlighted by the black dashed arrows. (c) The combination plot of intensity and spectra of -d$^3I$/d$V^3$ spectra in (b). (d) Typical -d$^3I$/d$V^3$ spectra at AA, AB, DW stacked regions from (c). (e) Evolution of the SC gap 2$\Delta$ along the same direction (blue dots). The red dashed line is the fitted curve to trace the oscillating gap. The details could be found in the main text and Supplementary Information. (f) Variation of the SC gap 2$\Delta$ for AA and AB stacked regions from Fig. 2a with $v$ ranging from -2.7 to -2.3.

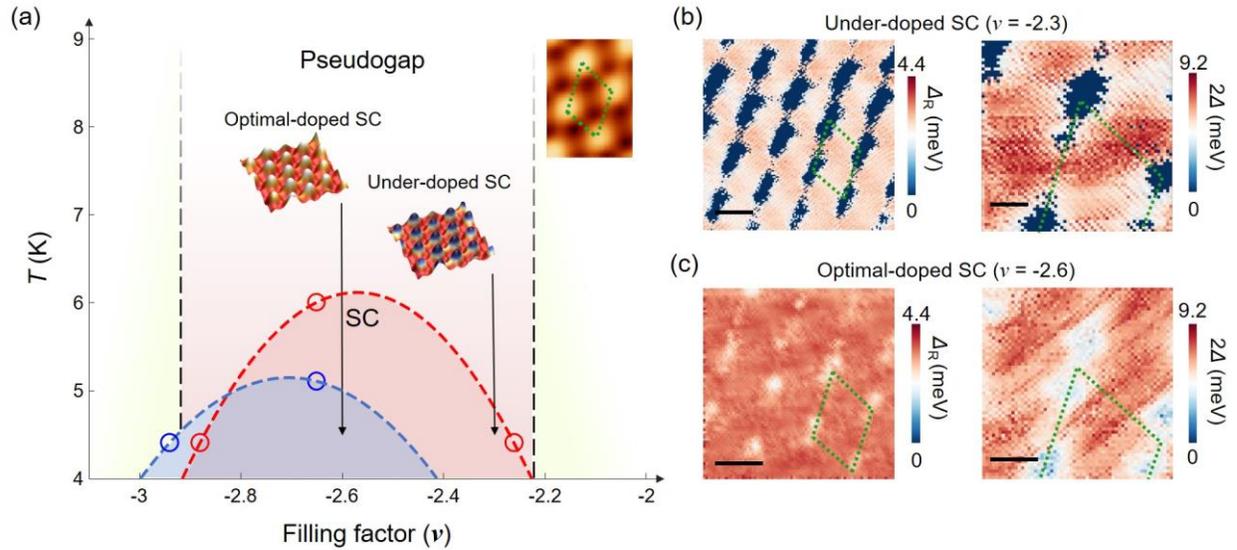

**Fig. 4 Directly visualizing the SC gap modulation.** (a) A schematic phase diagram for Sample 1 as a function of filling factor $v$ and temperature $T$ near SC regime. Blue shade represents SC in the AA stacked regions and red one represents SC in the AB stacked regions. The SC boundaries of AA and AB stacked region are qualitatively given by quadratic function fitting based on the data of PCS and temperature dependent spectra. Both AA and AB stacked regions show SC phase at optimal-doped regime with modulated gap size. At under-doped regime, only AB stacked regions shows SC phase. The inset show part of the STM topography (Large scale topography is in SI, Section XVI) for (b) and (c). The green dashed diamond represents a moiré unit cell, serving as a marker for consistent labeling the same position across panels (b) and (c). (b) The map of extracted right coherence peak energy $\Delta_R$ values (left; scale bar, 10 nm) and entire SC gap $2\Delta$ (right; scale bar, 4 nm) in under-doped regime ($v = -2.3$). (c) Same as (b) but for optimal-doped regime ($v = -2.6$).